# Using electrical impedance spectroscopy to identify equivalent circuit models of lubricated contacts with complex geometry: in-situ application to mini traction machine

Min Yu[a]✉, Jie Zhang[a], Arndt Joedicke[b], Tom Reddyhoff[a]

[a] *Department of Mechanical Engineering, Imperial College London, SW7 2AZ, London, United Kingdom*

[b] *Shell Technology Centre Hamburg, Hohe-Schaar-Str. 36, D-21107, Hamburg, Germany*

✉Corresponding author: m.yu14@imperial.ac.uk

**Abstract**

Electrical contact resistance or capacitance as measured between a lubricated contact has been used in tribometers, partially reflecting the lubrication condition. In contrast, the electrical impedance provides rich information of magnitude and phase, which can be interpreted using equivalent circuit models, enabling more comprehensive measurements, including the variation of lubricant film thickness and the asperity (metal-to-metal) contact area. An accurate circuit model of the lubricated contact is critical as needed for the electrical impedance analysis. However, existing circuit models are hand derived and suited to interfaces with simple geometry, such as parallel plates, concentric and eccentric cylinders. Circuit model identification of lubricated contacts with complex geometry is challenging. This work takes the ball-on-disc lubricated contact in a Mini Traction Machine (MTM) as an example, where screws on the ball, grooves on the disc, and contact close to the disc edge make the overall interface geometry complicated. The electrical impedance spectroscopy (EIS) is used to capture its frequency response, with a group of load, speed, and temperature varied and tested separately. The results enable an identification of equivalent circuit models by fitting parallel resistor-capacitor models, the dependence on the oil film thickness is further calibrated using a high-accuracy optical interferometry, which is operated under the same lubrication condition as in the MTM. Overall, the proposed method is applicable to general lubricated interfaces for the identification of equivalent circuit models, which in turn facilitates in-situ tribo-contacts with electric impedance measurement of oil film thickness – it does not need transparent materials as optical techniques do, or structural modifications for piezoelectric sensor mounting as ultrasound techniques do.

**Keywords:** electrical impedance; model identification; lubrication condition monitoring; film thickness.

## 1. Introduction

Lubrication condition monitoring is important for the maintenance of mechanical transmissions in transportation, power generation and other industrial equipment, with a huge market worth USD 1.0 billion in 2021 and estimated 1.4 billion in 2026 [1]. A group of non-destructive sensing techniques, including ultrasonic reflection [2]-[4], acoustic emission [5], optical interference [6]-[8], and laser induced fluorescence [9]-[12], have been employed to probe the lubrication performance related variables, such as oil film thickness, fluid viscosity, oil contamination, friction and so on. However, optical methods require transparent materials and are suited to in-lab research and high-accuracy film thickness calibration; acoustic methods need additional modifications in transmission bearings to accommodate piezoelectric sensors – this is sometimes not realistic for compact rotary machines, and therefore hindering in-field implementation of lubrication condition monitoring. In contrast, electrical methods measure resistance / capacitance / impedance between a tribo contact by simply applying two electrodes. For example, a function module known as the Electric Contact Resistance (ECR) has been deployed in PCS Instruments of Mini Traction Machine (MTM) and High Frequency Reciprocate Rig (HFRR) to the probe the variation of the resistance of ball-on-disc lubricated contacts, which can basically reflect the lubrication condition [13]-[15]. This is enabled by applying a DC voltage of 15 mV with the output information of the voltage across the ball-on-disc contact: a full-film lubrication is equivalent to an "open circuit" with its voltage being 100% of the applied 15 mV, a boundary lubrication with metal-to-metal contact means a "short circuit" with its voltage being 0% of the applied 15 mV, while a mixed boundary lubrication or a boundary lubrication with growing polymeric film (*e.g.*, ZDDP) may give a percentage between 0% and 100%. The capacitance measurement has been proven effective to probe the variation of the oil film thickness or the water content.

The electrical impedance spectroscopy (EIS) technique, which provides rich information of impedance magnitude/phase spectrum, has been widely used in biosensing, coating and battery [16]. More recently, the EIS has also shown a potential in the application to lubrication condition monitoring, including the variation of lubricant film thickness as estimated by the electrical capacitance; the transition between full-film, mixed, and boundary lubrication regimes, as differentiated using the electrical resistance; the forming of anti-wear boundary film, where an additional equivalent resistor in parallel to a capacitor are introduced; and the degradation of lubricant, such as fuel dilution, oil oxidization, and water emulsifying, which affect the values of resistance or capacitance in the equivalent electrical circuits by different extents [17]-[25]. An equivalent circuit model of a lubricated contact is always needed to interpret the change in these variables of interest. However, existing circuit models are hand derived and suited only to interfaces with simple geometry, such as parallel plates, concentric and eccentric cylinders. Circuit model identification of lubricated contacts with complex geometry are more challenging. This work takes a lubricated ball-on-disc contact in a tribometer of Mini Traction Machine (MTM, supplied by the PCS Instrument) as an example, where screws on the ball, grooves on the disc, and contact close to the disc edge make the overall interface geometry complicated. The electrical impedance spectroscopy (EIS) is used to capture its frequency response under steady-state condition, with a group of load, speed, and temperature tested respectively. The results enable an identification of equivalent circuit models that consist of resistors and capacitors, the dependence of which on the oil film thickness is further calibrated using high-accuracy optical interferometry (under the same lubrication condition as in the MTM). Overall, the proposed method is applicable to general lubricated interfaces for the identification of equivalent circuit models, which in turn facilitates in-situ tribo-contacts with the electric impedance measurement of oil film thickness – it does not need transparent materials as optical techniques do, or structural modifications for piezoelectric sensor mounting as ultrasound techniques do.

The main contributions of the present work include: i) electrical impedance spectroscopy (EIS) enabled identification of equivalent circuit models of tribo contacts with complex geometry, as demonstrated by mini traction machine (MTM) ball-on-disc contacts, and ii) development of a calculation model that converts the identified resistance and capacitance values to the central oil film thickness. The rest of this paper is organized as follows. Section 2 presents a group of equivalent circuit models that are typically used to characterize lubrication condition. Section 3 takes a ball-on-disc contact in a MTM as an example to detail the general procedures of using EIS to identify equivalent circuit models, which are further used to calculate film thickness with the calibration of high-accuracy optical interferometry. Section 4 discusses the independency of circuit models on lubricant temperature and categories, and also indicates EIS potential and challenges in both ex-situ and in-situ application to lubrication condition monitoring. Section 5 draws the conclusions.

## 2. Equivalent circuit models of tribo-contacts

### 2.1. Basic models and representation of Bode/Nyquist plots

Lubricated interfaces exhibit electric circuit characteristics by combining metal-to-metal asperity contact and electrical properties of lubricant itself, is illustrated in Figure 1. A group of equivalent circuit models that consist of resistors and capacitors have been mathematically given, including: i) the RC model, which is a resistor in parallel with a capacitor, ii) the RC circuit in series with another resistor, and iii) the Wagner impedance, which captures the diffusion process of lubricant. Moreover, hand derived models based on interface geometry have been proposed, such as the concentric / eccentric interfaces and the ball-on-disc contact. To better interpret and analyze electric circuit models, two representations of diagrams are normally used, including i) Bode diagram, which plots frequency response of both magnitude and phase of electrical impedance, and ii) Nyquist diagram, which plots the variation of both the real and imaginary parts of electrical impedance in a complex plane.

Both the Bode and Nyquist diagrams of the four typical circuit models (the RC model, the RC model in parallel with another resistor, and these models with the Warburg impedance taken into account) that describe the tribo-contacts are plotted in Figure 2 and detailed as follows.

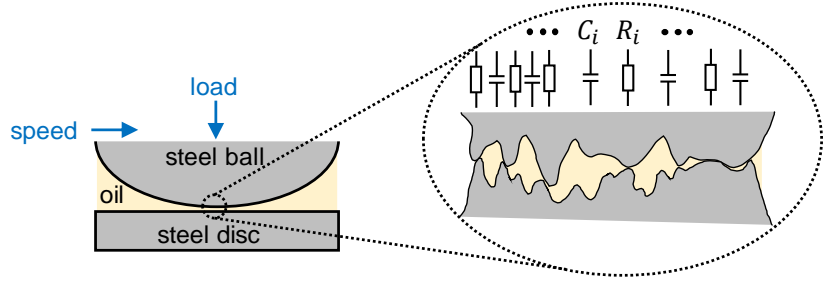

Figure 1: Illustration of parallel resistors ($R_i$) and capacitors ($C_i$) in a steel ball-on-steel disc contact under mixed lubrication regime.

The electrical impedance of resistor, capacitor and inductor are given below respectively:

$$Z_R = R \tag{1}$$

$$Z_C = \frac{1}{j\omega C} \tag{2}$$

$$Z_L = j\omega L \tag{3}$$

The electrical impedance of a resistor in parallel with a capacitor (*i.e.*, the RC model) is:

$$\frac{1}{Z_1} = \frac{1}{Z_{R_1}} + \frac{1}{Z_{C_1}} = \frac{1 + j\omega R_1 C_1}{R_1} \xrightarrow{yields} Z_1 = \frac{R_1}{1 + j\omega R_1 C_1} \tag{4}$$

Where the cut-off frequency (-3 dB) is $f_c = 1/R_1 C_1$ (or $\omega_c = 2\pi/R_1 C_1$). The Bode magnitude and Nyquist plots are provided in Figure 2-a). If another resistor ($R_2$) is introduced, then the overall electrical impedance is given below and plotted in Figure 2-b).

$$Z_2 = \frac{R_1}{1 + j\omega R_1 C_1} + R_2 = \frac{R_1 + R_2 + j\omega R_1 R_2 C_1}{1 + j\omega R_1 C_1} \tag{5}$$

The Warburg impedance (also referred to as diffusion element) models the diffusion process in dielectric spectroscopy. The Warburg impedance contains two elements in parallel: a diffusional capacitance and a diffusional resistance, each of them is proportional to the square root of the angular frequency ($\omega$), mathematically given in equation (6), where $A_W$ is the Warburg coefficient (or Warburg constant). The presence of the Warburg impedance can be recognized as a linear relationship exists with a slope of -1/2 in the low-frequency range of Bode plot ($\log|Z|$ versus $\log \omega$) or as a linear relation angled at 45° in the low-frequency section of Nyquist plot (see in Figure 2-c) and -d)).

$$Z_W = \frac{A_W}{\sqrt{\omega}} + \frac{A_W}{j\sqrt{\omega}} \tag{6}$$

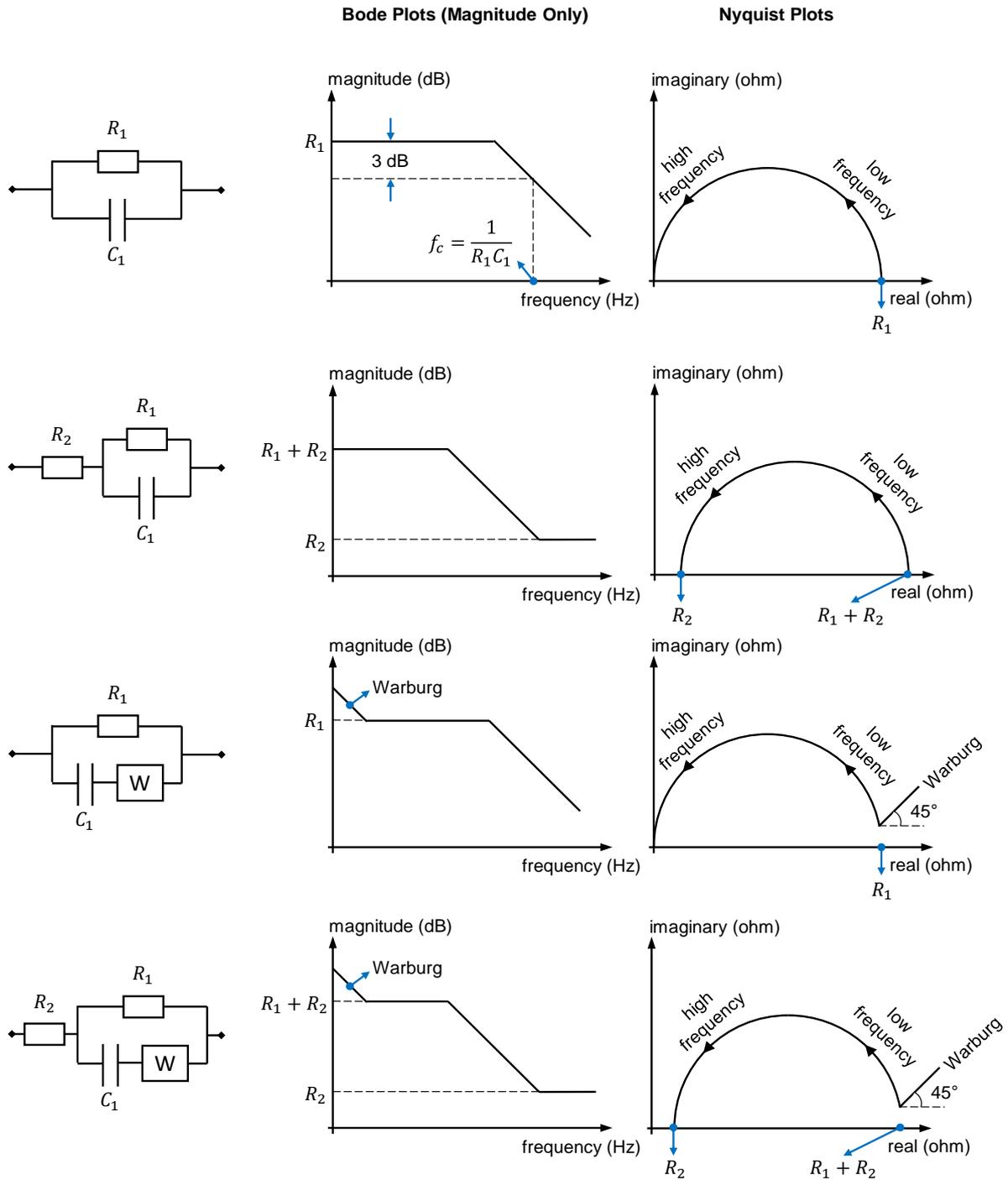

Figure 2: Equivalent circuit models of tribo-contacts and their representations in Bode (magnitude) and Nyquist plots.

### 2.2. Hand-derived circuit model of tribo-contacts

Maruyama & Nakano have derived an equivalent circuit model of a lubricated ball-on-disc contact that essentially captures the variation of the resistance and the capacitance from full-film, mixed to boundary lubrication regimes [26]. As illustrated in Figure 3-a), the basic assumptions are stated as follows. The overall electric circuit is built by the deformed Hertz contact area and its surrounding area, some metal-to-metal asperity contacts (named as "electric breakdown", also illustrated in Figure 1) happens within the Hertz contact area, thus the Hertz area can be divided into two parts: "electric breakdown" and "non electric breakdown" areas. The circuit resistance is determined only by the "electric breakdown" area within the Hertz contact, while the circuit capacitor is made jointly by combining the "non electric breakdown" area within the Hertz contact and the surrounding area. It is understood that this electrical breakdown area will be increasing, and the central film thickness will be decreasing with larger load, lower speed, higher temperature.

According to the geometrical model derived in [26], the capacitance of the "non electric breakdown" area within the Hertz contact can be approximated by a parallel-plate capacitor:

$$C_1 = \varepsilon \pi r^2 (1 - \alpha)/h_c \tag{7}$$

where $r$ is the radius of Hertz contact area, and $\alpha$ is the breakdown ratio (*i.e.*, the electric breakdown area divided by the whole Hertz contact area).

The capacitance of the surround area is calculated as equation (8) below, according to [26]:

$$C_2 = 2\pi\varepsilon \left(h_c + \sqrt{r^2 - a^2}\right) \left[\ln\left(\frac{r}{h_c + \sqrt{r^2 - a^2}}\right) - 1\right] \tag{8}$$

The overall capacitance of the ball-on-disc lubricated contact is $C_1 + C_2$. The resistance of the ball-on-disc lubricated contact is related only to the electric breakdown area within the Hertz contact:

$$R_1 = R_0/\alpha \tag{9}$$

$R_0$ is the contact resistance under the stationary contact condition, where the electric breakdown area equals the Hertz contact area of $\pi r^2$.

In terms of a lubricated cylindrical contact, such as in a journal bearing, the capacitance is given as below [27]:

$$C_3 = \frac{2\pi\varepsilon L}{\text{arcosh}\left(\frac{r_1^2 + r_2^2 - d^2}{2 r_1 r_2}\right)} \tag{10}$$

where the associated parameters are referred to Figure 3-b), and $L$ is length of the cylindrical contact. Similarly, the resistance of the lubricated cylindrical contact, $R_3$, is determined by the electric breakdown area.

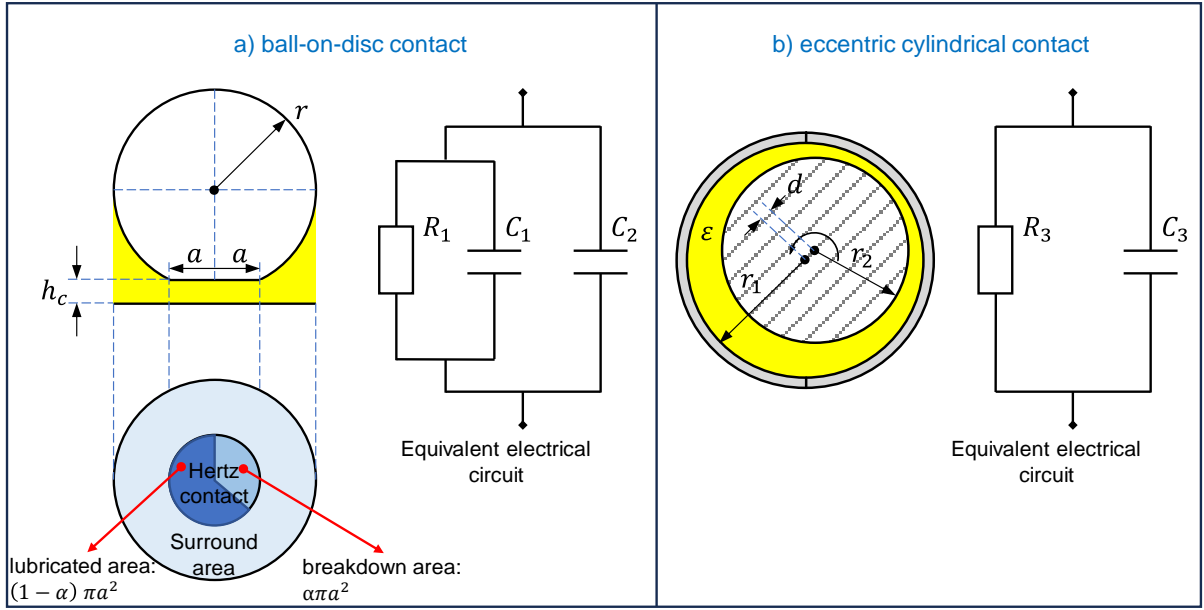

Figure 3: Hand-derived equivalent circuit models of typical lubricated tribo-pairs: a) a ball-on-disc contact (such as in a roller bearing), proposed by Maruyama & Nakano [26], and b) an eccentric cylindrical contact (such as in a journal bearing), given by Dawes [27].

## 3. Electrical impedance spectrum identified equivalent circuit model

The above hand-derived electric circuit models are suited to tribo-contacts with simple geometry, however, in many cases the lubricated interfaces are with complex geometry, making electrical impedance method difficult to be used for lubrication condition monitoring. To address this issue, This work takes a ball-on-disc contact in a Mini Traction Machine (MTM) as an example, where screws on the ball, grooves on the disc, and contact close to the disc edge make the overall interface geometry irregular (as illustrated in Figure 4-a)), making the overall circuit model deviate from that of an ideal ball-on-disc. A novel method that employs electrical impedance spectrum (EIS) to identify equivalent circuit models is proposed and detailed in the following steps from a) to e). This method can be also generalized to deal with other lubricated contacts with complex geometry.

**Step a):** As shown in Figure 4-a), a potentiostat (Interface 1010E, supplied by Gamry Instruments) is adopted to implement the measurement of electrical impedance spectrum (EIS), where the frequency of the applied sinusoidal voltage is capable of being swept from 10 µHz to 2 MHz. The two electrodes of the potentiostat are connected to the ball and the disc respectively in a MTM (practically connected to the ball and disc wires in the MTM that are originally designed for the built-in electric circuit resistance module). The EIS measurement from 1 MHz to 1 Hz normally takes 3 mins, therefore during this period the ball-on-disc is forced under steady-state operation, with constant oil temperature ($T$), constant load ($W$), and constant entrainment sliding speed ($U$), as indicated in Figure 4-a). Importantly, the amplitude of the input sinusoidal signal is 10 mV, this is to avoid any influence from the electrical potential on the lubrication itself [28]. An example of EIS measurement with *T* = 60C, *W* = 20N, and *U* = 2500 mm/s is plotted in the Bode diagram in Figure 4-a) -right, the corresponding resistance (*R*) and capacitance (*C*) of the circuit can be identified by fitting the EIS data to a simple RC circuit model (a resistor in parallel with a capacitor). It is worth noting that higher order circuit models are not adopted to identify R and C values, this is because higher order circuit models may also incorporate the undesirable EIS data noise.

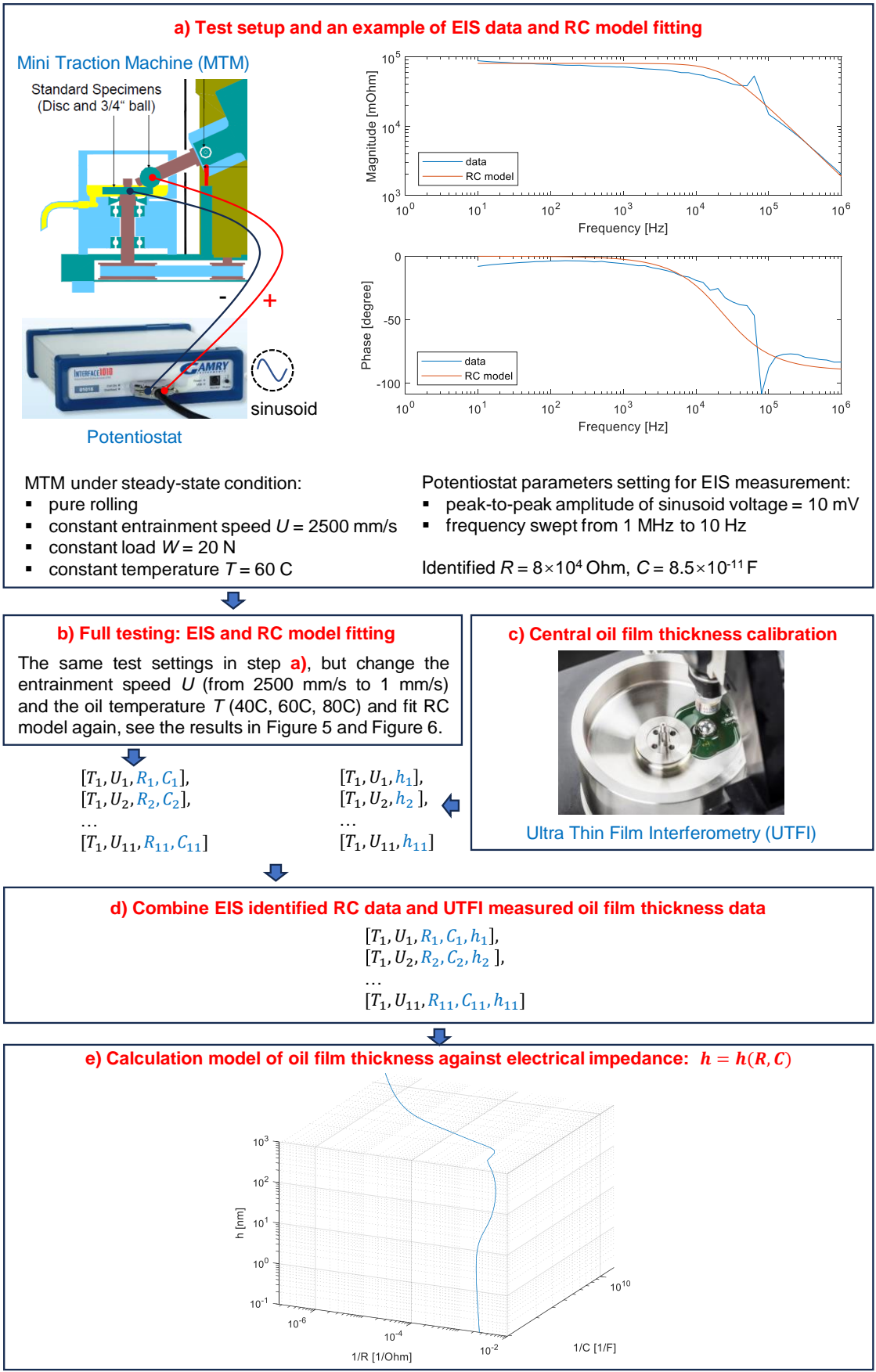

Figure 4: Procedures of identifying the oil film thickness calculation model ($h = h(R, C)$) for the lubricated ball-on-disc contact in a Mini Traction Machine (MTM).

**Step b):** Full EIS testing is further performed with varied entrainment speed *U* = [2500, 2000, 1300, 1000, 700, 500, 200, 100, 50, 10, 1] mm/s, and the EIS data is provided in Bode diagrams in Figure 5, where the R and C values for each step of *U* can be identified by following the same method in the step a) and also plotted in Figure 6. It can be seen that from high speed to low-speed operation, the lubrication gradually moves from hydrodynamic lubrication regimes (in the cases of *U* = [2500, 2000, 1300, 1000, 700, 500] mm/s, where the capacitance increases as the film thickness reduces step by step, reflected by the zoom-in view) to the boundary lubrication (where short circuit is clear with low DC gains, around 10 Ohms). Interestingly the mixed lubrication (in the case around *U* = 200 mm/s) with intermittent metal-to-metal contact leads to more significant fluctuation in the phase angle of the EIS (see in the bottom plot in Figure 5), accompanying by mechanical and electrical noise.

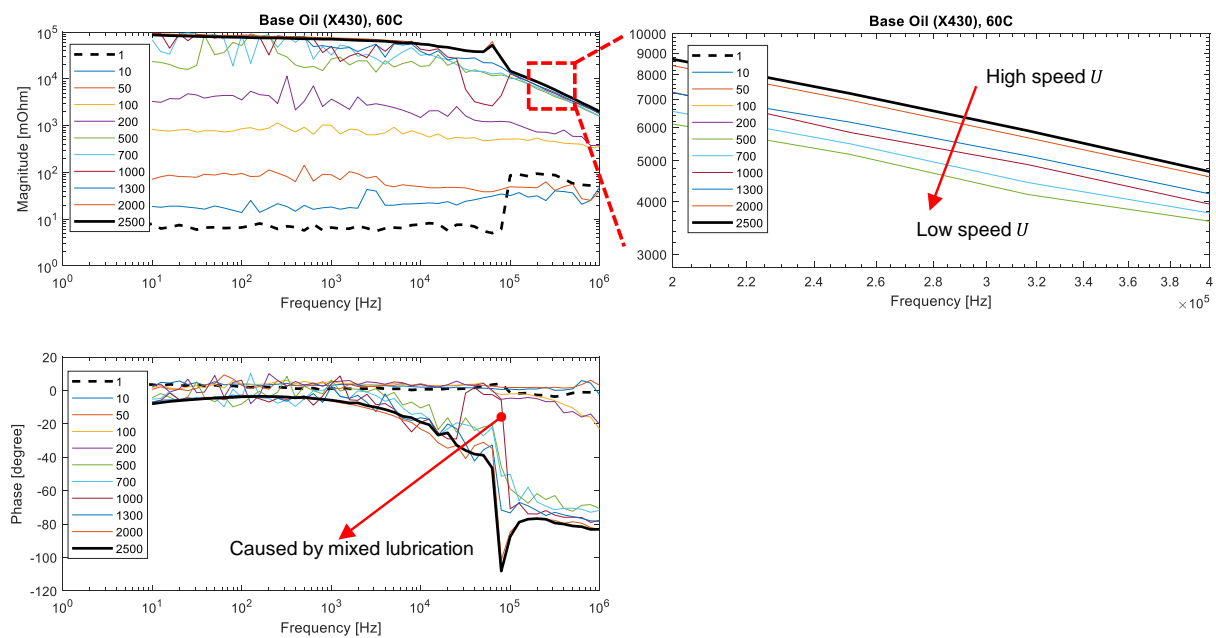

Figure 5: Electrical impedance magnitude (and its zoom-in view) and phase spectrum of a lubricated ball-on-disc contact in the MTM under steady-state condition, where a group of entrainment speed ranging from 2500 mm/s to 1 mm/s as listed inside the legend box are tested separately. The numbers in the legend box refer to the MTM entrainment speed *U*.

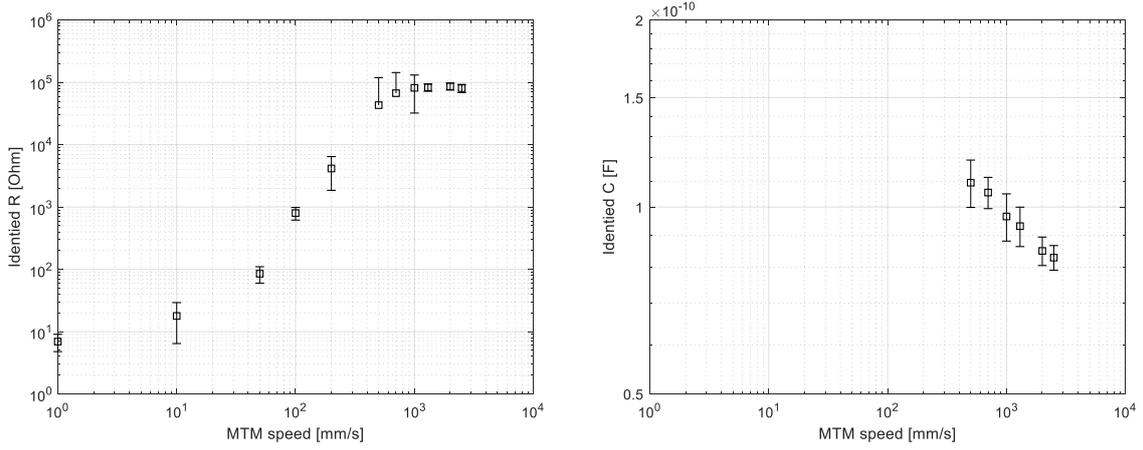

Figure 6: EIS data Identified R and C values with varied MTM entrainment speed.

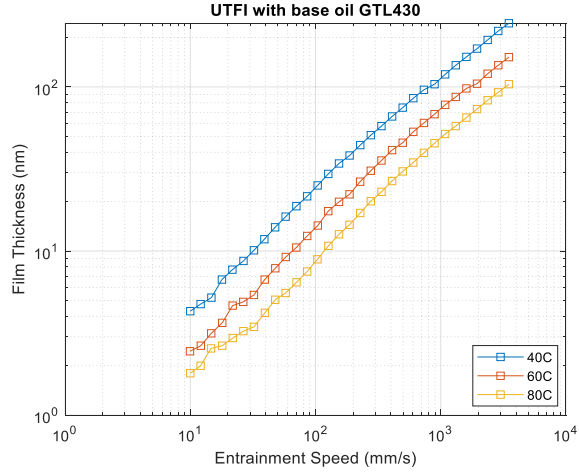

Figure 7: Ultra-Thin Film Interferometer (UTFI) calibrated central film thickness (*h*) under elastohydrodynamic (EHD) lubrication. The same oil (base oil GTL430), the same load (20 N), the same temperature (40C, 60C, and 80C), and the same pure rolling contact as in the MTM testing are adopted.

**Step c)**: To calibrate the oil film thickness, Ultra-Thin Film Interferometer (UTFI, supplied by PCS Instruments [7]) is employed to measure the actual value of the film thickness under the elastohydrodynamic (EHD) lubrication regime. To compensate for the effect of steel ball-on-glass disc in UTFI, the measured central film thickness has been adjusted to adapt to the MTM test setup of steel ball-on-steel disc.

The Dowson and Hamrock equation is given below:

$$h_c = kR' \left(\frac{U\eta_o}{E'R'}\right)^{0.68} (\alpha E')^{0.53} \left(\frac{W}{E'R'^2}\right)^{-0.067} \tag{11}$$

where $k$ is determined by the geometry of elliptical contact area, $E'$ and $R'$ are the reduced Young's modulus and reduced radius of the interacting solids, $W$ is the applied load, $U$ is the entrainment speed of the ball-on-disc contact, $\eta_o$ is the dynamic viscosity at atmosphere pressure, and $\alpha$ is the pressure viscosity coefficient.

Convert the EHD central film thickness in a steel ball-on-glass disc contact as measured in the UTFI ($h_{c\_UTFI}$) to that in a steel ball-on-steel disc contact as in the MTM ($h_{c\_MTM}$),

$$\frac{h_{c\_MTM}}{h_{c\_UTFI}} = \frac{kR'\left(\frac{U\eta_o}{E'_{ss}R'}\right)^{0.68}(\alpha E'_{ss})^{0.53}\left(\frac{W}{E'_{ss}R'^2}\right)^{-0.067}}{kR'\left(\frac{U\eta_o}{E'_{sg}R'}\right)^{0.68}(\alpha E'_{sg})^{0.53}\left(\frac{W}{E'_{sg}R'^2}\right)^{-0.067}} = \left(\frac{E'_{ss}}{E'_{sg}}\right)^{-0.083} \quad (12)$$

where the reduced Young's Modulus for a steel ball-on-glass disc and a steel ball-on-steel disc are $E'_{sg} = 1.17 \times 10^{11}$ N/m² and $E'_{ss} = 2.26 \times 10^{11}$ N/m² respectively. Therefore the film thickness conversion can be given as: $h_{c\_MTM} = 0.95 \cdot h_{c\_UTFI}$.

**Step d)**: After the R and C in the dataset $[T_i, U_j, R_j, C_j]$ identified from step b) and the central oil film thickness $h$ in the dataset $[T_i, U_j, h_j]$ calibrated from step c), a new dataset $[T_i, U_j, R_j, C_j, h_j]$ can be formulated.

**Step d)**: With $[U_j, R_j, C_j, h_j]$ in the newly formulated dataset at three different temperature of $T_1$ = 40 C, $T_2$ = 60 C, and $T_3$ = 80 C, the oil film thickness calculation model with electrical impedance, $h = h(R, C)$, can be eventually identified and shown in Figure 8. Interestingly, all data points $[R_j, C_j, h_j]$ collapse in a single curve irrespective of the temperature values.

With the identified circuit models, the frequency response of the lubricated MTM ball-on-disc contact can be plotted in the representation of Bode Diagram and Nyquist Diagram in Figure 9, where the central film thickness $h$ mathematically varied from $10^{-1}$ nm to $10^3$ nm. This can be more directly compared to the raw EIS data measured by the potentiostat (which are presented in Bode / Nyquist diagrams), and then used to estimate the lubrication condition.

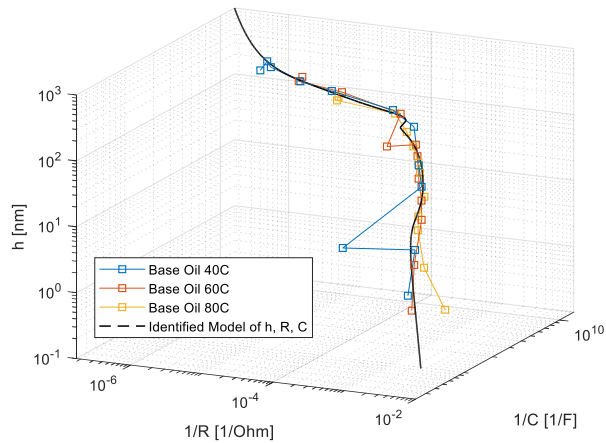

Figure 8: Identified model of the central film thickness (of the lubricated ball-on-disc contact in the MTM) with respect to the capacitance and the resistance.

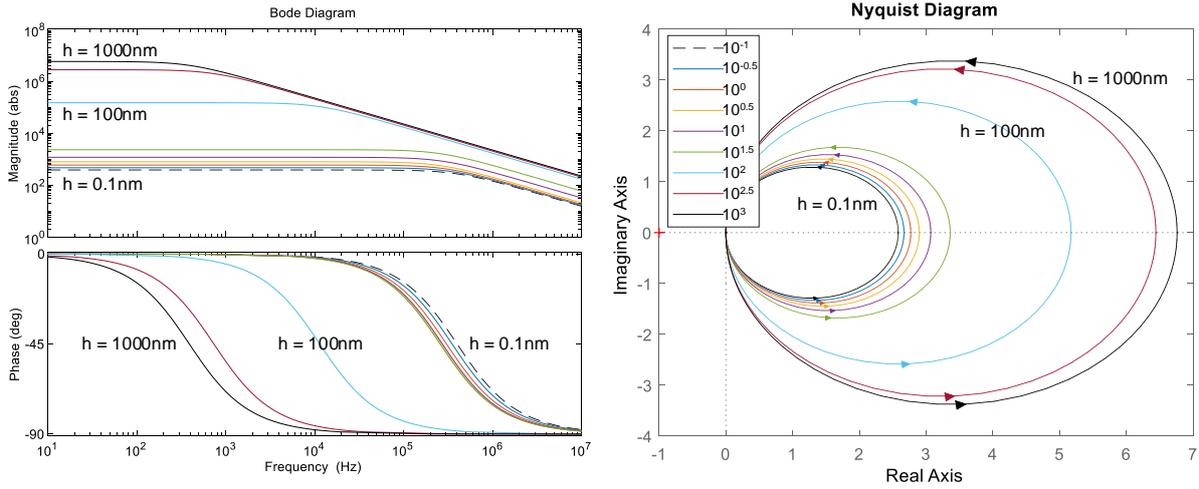

Figure 9: Frequency response of the identified circuit models of the lubricated MTM ball-on-disc contacts in the representation of Bode Diagram and Nyquist Diagram (the legend refers to the central oil film thickness $h$ in nanometres), where $h$ is mathematically ranged from $10^{-1}$ nm to $10^3$ nm.

## 4. Discussion

According to the Dowson and Hamrock equation in (11), the MTM ball-on-disc central film thickness under the EHD lubrication regime is determined by oil viscosity, the entrainment speed, and the load, that is $h = h(\eta_o, U, W)$. In contrast, according to the electrical impedance measured oil film thickness equation, $h = h(R, C)$, as identified in Section 3, if two MTM contacts have the same oil thickness, then it means they also the same resistance and capacitance, and vice versa – this is independent on oil category and the variation of oil temperature. A detailed explanation is given as follows. Figure 10 provides the variation of dielectric constants ($\varepsilon_r$) of different fluids (fully formulated oils of 0W-30 and 0W-16, base oils of GTL420 and GTL430, and diesel) against the temperature, the changes in dielectric constants are highly limited (less than 3%) particularly within the MTM operation temperature range. However, if the lubricant is diluted by water, which significantly affects fluid dielectric constant and thus the capacitance value, then the identified model $h = h(R, C)$ will not stand and has to be modified. On the other hand, the changes in the electrical resistance of the lubricant itself (due to varied temperature, fuel contamination, aging, etc.) will not affect the identified model $h = h(R, C)$, this is because the conductivity of the stainless steel is much higher than that of the lubricant (see in Table 1), the metal-to-metal asperity contact (*i.e.*, the electric breakdown) always dominates the resistance value of the overall contact irrespective of the lubricant resistance level.

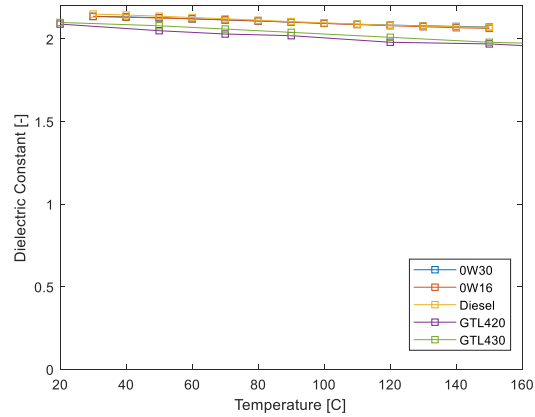

Figure 10: Variation of dielectric constants ($\varepsilon_r$) of different fluids (fully formulated oils of 0W-30 and 0W-16, base oils of GTL420 and GTL430, and diesel) against the temperature.

Table 1 conductivity of different materials

|  | Stainless steel | Lubricant | Glass | Air |
|---|---|---|---|---|
| Conductivity (S/m) | 1.5×10$^6$ | 10$^{-12}$ | 10$^{-11}$ ~ 10$^{-15}$ | 10$^{-9}$ ~ 10$^{-15}$ |

The present work proposes a novel method that uses the electrical impedance spectroscopy (EIS) to identify electric circuit models of lubricated contacts for oil thickness measurement, this model is independent on the variation of oil temperature, oil categories, and cavitation, and also has advantages of no transparent materials needed as optical techniques do, no structural modifications needed for sensors mounting as ultrasound techniques do. However, the following points remain challenging and will be addressed in the future work.

- The present experimental testing with MTM uses a lubricant specimen of base oil (Shell GTL430), which has no polymeric additives. However, in case of a fully formulated oil, the slowly growing boundary film (e.g., ZDDP) will introduce another resistor-capacitor, making the overall electric circuit of the dynamic lubricated contact more complicated.

- Signal noise in the EIS magnitude and phase is the main factor that affect the accuracy the identified model and thus the calculation of the oil film thickness. The noise may be mechanical, such as tilt of the MTM disc during the rotation – this can be possibly addressed by triggering the electrical impedance measurement at fixed angular position of the disc, or electrical, such as other inductors and capacitors between the MTM ball and disc – this needs to be electrically isolated by wiring to ground or using shields.

- Any water emulsified in the oil specimen will significantly change the dielectric constant of lubricant and thus deviates the identified model. This however will possibly enable additional measurement of water content in oil exactly located between the ball and disc, this is particularly beneficial for water-in-oil emulsion related research project.

In addition to the electrical impedance spectroscopy (EIS) identified electric circuit models of lubricated contacts for oil film thickness measurement as proposed in this work, the EIS also has a promising potential in both ex-situ and in-situ application to lubrication condition monitoring.

- Ex-situ application: the EIS can be used to detect the degradation and contamination of lubricant specimens, including fuel dilution, oil aging, and water emulsifying, which affect the values of resistance or capacitance of lubricant by different extents.

- In-situ application: the EIS impedance measured magnitude and phase spectrum can be used to identify the equivalent circuit models of lubricated contacts with complex interface geometry, for example, the line contact in a roller-raceway bearing. Additionally, with the identified model, the EIS can enable more comprehensive lubrication condition monitoring, in terms of the oil film thickness, the electrical breakdown area (*i.e.*, the asperity metal-to-metal contact area), and possibly the boundary film growth.

## 5. Conclusions

This work proposes a novel method that uses the electrical impedance spectroscopy (EIS) to identify equivalent circuit models of lubricated contacts with complex interface geometry. Particularly, the ball-on-disc in a Mini Traction Machine (MTM) is taken as example, where screws on the ball, grooves on the disc, and contact close to the disc edge make the circuit model difficult to be hand derived. The EIS is employed to capture the frequency response of the MTM ball-on-disc contact under steady-state condition, with a group of entrainment speed and temperature varied and tested separately. The results enable an identification of equivalent circuit models that consist of resistors and capacitors. Aided by a high-accuracy optical interferometry, the MTM ball-on-disc oil film thickness can be calibrated and eventually calculated only with electrical resistance and capacitance values. Importantly, this model is insensitive to the variation of oil electrical properties. This is because, on one hand, the changes in dielectric constants are highly limited (less than 3%) particularly within the MTM operation temperature range, also between different oil categories; on the other hand, the conductivity of the stainless steel is much higher than that of the lubricant, and therefore metal-to-metal asperity contact always dominates the resistance value of the overall contact irrespective of the lubricant resistance level.

Overall, the EIS enables identification of equivalent circuit models with complex interface geometry, in contrast, hand derived circuit models are only applicable to simple geometry (parallel plates, ideal ball-on-disc, concentric / eccentric cylindrical contacts, etc.). This method further facilitates existing tribometers with extra measurement of oil film thickness, while it does not need transparent materials as optical techniques do, or structural modifications for piezoelectric sensors mounting as ultrasound techniques do.

## Acknowledgements

The authors acknowledge the support from Taiho Kogyo Tribology Research Foundation (21B09).